\documentclass[doublecol]{epl2}
\usepackage{amsmath}
\usepackage{amssymb}
\usepackage{epsfig}
\usepackage{color}
\usepackage{subfigure}
\usepackage{graphicx,amsmath}
\usepackage{graphicx}

\title{Heat transport in magnetic fields by quantum spin liquid in the organic
insulators  $\rm \bf EtMe_3Sb[Pd(dmit)_2]_2$ and $\rm \bf
\kappa-(BEDT-TTF)_2Cu_2(CN)_3$} \shorttitle{Heat transport in
magnetic fields by quantum spin liquid}
\author{V. R. Shaginyan \inst{1,2}\thanks {Email:
\email{vrshag@thd.pnpi.spb.ru}} \and A. Z. Msezane \inst{2}\and K.
G. Popov\inst{3} \and G. S. Japaridze \inst{2} \and V. A. Khodel
\inst{4,5}} \shortauthor{V.R. Shaginyan \etal} \institute{\inst{1}
Petersburg Nuclear Physics Institute, Gatchina, 188300,
Russia\\\inst{2} Clark Atlanta University, Atlanta, GA 30314, USA\\
\inst{3} Komi Science Center, Ural Division, RAS, Syktyvkar,
167982, Russia\\ \inst{4} Russian
Research Center Kurchatov Institute, Moscow, 123182, Russia\\
\inst{5} McDonnell Center for the Space Sciences \& Department of
Physics, Washington University, St. Louis, MO 63130, USA}

 \pacs{66.}{Nonelectronic
transport properties of condensed matter}

\abstract{Measurements of the low-temperature thermal conductivity
collected on insulators with geometrical frustration produce
important experimental facts shedding light on the nature of
quantum spin liquid composed of spinons. We employ a model of
strongly correlated quantum spin liquid located near the fermion
condensation phase transition to analyze the exciting measurements
of the low-temperature thermal conductivity in magnetic fields
collected on the organic insulators $\rm EtMe_3Sb[Pd(dmit)_2]_2$
and $\rm \kappa-(BEDT-TTF)_2Cu_2(CN)_3$. Our analysis of the
conductivity allows us to reveal a strong dependence of the
effective mass of spinons on magnetic fields, to detect a scaling
behavior of the conductivity, and to relate it to both the
spin-lattice relaxation rate and the magnetoresistivity. Our
calculations and observations are in a good agreement with
experimental data.}

\begin{document}
\maketitle

The organic insulators $\rm EtMe_3Sb[Pd(dmit)_2]_2$ and $\rm
\kappa-(BEDT-TTF)_2Cu_2(CN)_3$ have two-dimensional triangular
lattices with the geometric frustration prohibiting the formation
of spin ordering even at the lowest accessible temperatures $T$
\cite{yam,yam1,yam2,scqsl,chqs1}. Therefore, these insulators offer
unique insights into the physics of quantum spin liquids (QSL).
Indeed, measurements of the heat capacity on the both insulators
reveal a $T$-linear term indicating that the low-energy excitation
spectrum from the ground state is gapless \cite{yam,yam1,yam2}. The
excitation spectrum can be deduced from measurements of the heat
conductivity $\kappa(T)$ in the low temperature regime. For
example, at $T\to0$ a residual value in $k/T$ signals that the
excitation spectrum is gapless. The presence of the residual value
is clearly resolved in $\rm EtMe_3Sb[Pd(dmit)_2]_2$, while
measurements of $k/T$ on $\rm \kappa-(BEDT-TTF)_2Cu_2(CN)_3$
suggests that the low-energy excitation spectrum can have a gap
\cite{scqsl,chqs1}. Taking into account the observed $T$-linear
term of the heat capacity in $\rm \kappa-(BEDT-TTF)_2Cu_2(CN)_3$
with the static spin susceptibility remaining finite down to the
lowest measured temperatures \cite{shim}, the presence of the spin
excitation gap becomes questionable. Thermal conductivity probe
elementary itinerant excitations and is totally insensitive to
localized ones such as those responsible for Schottky
contributions, which contaminates the heat capacity measurements at
low temperatures \cite{yam,yam1,yam2,scqsl,chqs1}. The heat
conductivity is formed primarily by both acoustic phonons and
itinerant spinons, while the latter form QSL. Since the phonon
contribution is insensitive to the applied magnetic field $B$, the
elementary excitations of QSL can be further explored by the
magnetic field dependence of $k$. Measurements under the
application of magnetic field $B$ of the heat conductivity $\kappa$
on these insulators have exhibited a strong dependence of
$\kappa(B,T)$ as a function of $B$ at fixed $T$ \cite{scqsl,chqs1}.
The obtained dependence at low temperatures resembles that of the
spin-lattice relaxation rate $(1/T_1T)$ at fixed temperature as a
function of magnetic field \cite{eplh}: $k(B)$ at low fields is
insensitive to $B$, displaying a response to increasing magnetic
field $B$. On the other hand, it is suggested that the observed
$B$-dependence implies that some spin-gap–like excitations coupling
to the magnetic field are also present at low temperatures
\cite{scqsl,chqs1}. As a result, we face a challenging problem of
interpretation of the experimental data in a consistent way,
including the $B$-dependence of the heat conductivity.

In this letter we explain the $B$-dependence of the low-temperature
thermal conductivity $\kappa$ in $\rm EtMe_3Sb[Pd(dmit)_2]_2$ and
$\rm \kappa-(BEDT-TTF)_2Cu_2(CN)_3$. We employ a model of strongly
correlated quantum spin liquid (SCQSL) located near the fermion
condensation phase transition (FCQPT) to analyze the $B$-dependence
of $\kappa$ \cite{pr,prbh,eplh,plah}. Our analysis allows us to
detect a scaling behavior of $\kappa(B,T)$, and relate it to the
scaling behavior of the spin-lattice relaxation rate $(1/T_1T)$
measured on both the herbertsmithite $\rm ZnCu_3(OH)_6Cl_2$ and the
heavy-fermion (HF) metal $\rm YbCu_{5-x}Au_{x}$, and the
magnetoresistivity measured on the HF metal $\rm YbRh_2Si_2$. Our
calculations are in a good agreement with experimental data.

\begin{figure}[!ht]
\begin{center}
\includegraphics [width=0.47\textwidth]{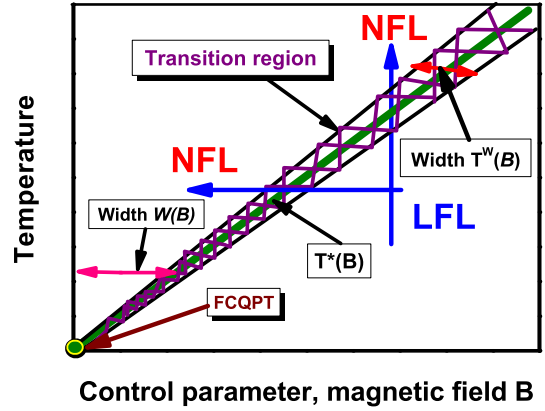}
\end{center}
\caption{(Color online). Schematic $T-B$ phase diagram of SCQSL
with magnetic field as the control parameter. The vertical and
horizontal arrows show LFL-NFL and NFL-LFL transitions at fixed $B$
and $T$, respectively. The hatched area represents the transition
region taking place at $T^*(B)$. The solid line in the hatched area
represents the function $T^*(B)\simeq T_M(B)$ given by
eq.~\eqref{TMB}. The functions $W(B)\propto T\propto T^*$ and
$T^W(B)\propto T\propto T^*$ shown by two-headed arrows define the
width of the NFL state and the transition area, respectively. At
FCQPT indicated by the arrow the effective mass $M^*$ diverges and
both $W(B)$ and $T^W(B)$ tend to zero.}\label{fig0}
\end{figure}

Representing a special case of QSL, SCQSL is a quantum state of
matter composed of spinons - chargeless fermionic spinons with spin
$1/2$ \cite{prbh,eplh,plah}. In insulating compounds, SCQSL can
emerge when interactions among the magnetic components  are
incompatible with the underlying crystal geometry, leading to a
geometric frustration generated by the triangular and kagome
lattices of magnetic moments, as it is in the case of $\rm
ZnCu_3(OH)_6Cl_2$, see e.g.
\cite{nath,bal,herb1,herb2,herb3,herb4,herb5}. In case of ideal
two-dimensional (2D) lattice the frustration of the lattice leads
to a dispersionless topologically protected branch of the spectrum
with zero excitation energy known as the flat band
\cite{green,vol,vol1}. Then, FCQPT can be considered as quantum
critical point of SCQSL, composed of chargeless heavy spinons with
$S=1/2$ and the effective mass $M^*_{mag}$, occupying the
corresponding Fermi sphere with the Fermi momentum $p_F$.
Therefore, the properties of insulating compounds coincide with
those of heavy-fermion metals with one exception: it resists the
flow of electric charge \cite{prbh,eplh,plah}. As we are dealing
with compounds confining non-ideal triangular and kagome lattices,
we have to bear in mind that the real magnetic interactions and
possible distortion of the lattices can shift the SCQSL from the
exact FCQPT, positioning it somewhere near FCQPT. Therefore, the
actual location of the SCQSL phase in fig. \ref{fig0} has to be
established by analyzing the experimental data only.

In the vicinity of the FCQPT, pronounced deviations from the
behavior of Landau Fermi liquid (LFL) with regard to physical
properties are observed. These so-called non-Fermi-liquid (NFL)
phenomena are related to the action of strong enhancement of the
effective mass $M^*_{mag}$ associated with the FCQPT. We note that
there are different kinds of instabilities of LFL connected with
several perturbations of initial quasiparticle spectrum
$\varepsilon({\bf p})$ and occupation numbers $n({\bf p})$,
associated with strong enhancement of the effective mass and
leading to the emergence of a multi-connected Fermi surface, see
e.g. \cite{pr,asp,zvbld,khodb}. Depending on the parameters and
analytical properties of the Landau interaction, such instabilities
lead to several possible types of restructuring of the initial LFL
ground state. This restructuring generates topologically distinct
phases. One of them is the fermion condensation associated with
FCQPT, another one belongs to a class of topological phase
transitions, where the sequence of rectangles $n(p)=0$ and $n(p)=1$
is realized at $T=0$. In fact, at elevated temperatures the systems
located at these transitions exhibit behavior typical to those
located at FCQPT \cite{pr}. Therefore, we do not consider the
specific properties of these topological transitions, and focus on
the behavior of systems located near FCQPT.

We start with a brief outline of the effective mass dependence on
magnetic field and temperature, $M^*_{mag}(B,T)$. The key point of
the formalism is the extended quasiparticle paradigm when the
effective mass is no more constant but depends on temperature $T$,
magnetic field $B$ and other external parameters such as pressure
$P$ \cite{pr}. To study the low temperature transport properties,
scaling behavior, and the effective mass $M^*_{mag}(B,T)$ of SCQSL,
we use the model of homogeneous HF liquid. In that case, the model
permits to avoid complications associated with the crystalline
anisotropy of solids \cite{pr}, while the Landau equation,
describing the effective mass $M^*$ of HF liquid, reads
\cite{land,pr}
\begin{eqnarray}
\nonumber \frac{1}{M^*_{\sigma}(B,
T)}&=&\frac{1}{M}+\sum_{\sigma_1}\int\frac{{\bf p}_F{\bf
p}}{p_F^3}F_
{\sigma,\sigma_1}({\bf p_F},{\bf p}) \\
&\times&\frac{\partial n_{\sigma_1} ({\bf
p},T,B)}{\partial{p}}\frac{d{\bf p}}{(2\pi)^3}, \label{HC1}
\end{eqnarray}
where $M$ is the corresponding bare mass, $F_{\sigma,\sigma_1}({\bf
p_F},{\bf p})$ is the Landau interaction, which depends on Fermi
momentum $p_F$, momentum $p$ and spin index $\sigma$. The
distribution function $n$ can be expressed as
\begin{equation}
n_{\sigma}({\bf p},T)=\left\{ 1+\exp \left[\frac{(\varepsilon({\bf
p},T)-\mu_{\sigma})}T\right]\right\} ^{-1},\label{HC2}
\end{equation}
where $\varepsilon({\bf p},T)$ is the single-particle spectrum. In
our case, the chemical potential $\mu$ depends on the spin due to
Zeeman splitting $\mu_{\sigma}=\mu\pm \mu_BB$, $\mu_B$ is Bohr
magneton.

In LFL theory, the single-particle spectrum is a variational
derivative of the system energy $E[n_{\sigma}({\bf p},T)]$ with
respect to occupation number $n$, $\varepsilon({\bf p},T)=\delta
E[n({\bf p})]/\delta n$. Choice of the interaction and its
parameters is dictated by the fact that the system has to be at
FCQPT \cite{pr,ckz,epl}. Thus, the sole role of the Landau
interaction is to bring the system to the FCQPT point, where Fermi
surface alters its topology so that the effective mass acquires
temperature and field dependence \cite{pr,ckz,khodb}. The
variational procedure, being applied to the functional
$E[n_{\sigma}({\bf p},T)]$, gives the following form for
$\varepsilon_\sigma({\bf p},T)$,
\begin{equation}\label{epta}
\frac{\partial\varepsilon_\sigma({\bf p},T)}{\partial{\bf p}}
=\frac{{\bf p}}{M}-\sum_{\sigma_1}\int \frac{\partial F_
{\sigma,\sigma_1}({\bf p},{\bf p}_1)}{\partial{\bf
p}}n_{\sigma_1}({\bf p}_1,T)\frac{d^3p_1}{(2\pi)^3},
\end{equation}

Equations \eqref{HC2} and \eqref{epta} constitute the closed set
for self-consistent determination of $\varepsilon_\sigma({\bf
p},T)$ and $n_{\sigma}({\bf p},T)$ and the effective mass,
$p_F/M^*_{mag}=\partial\varepsilon(p)/\partial(p)|_{p=p_F}$. We
emphasize here, that in our approach the entire temperature and
magnetic field dependence of the effective mass is brought to us by
dependencies of $\varepsilon_\sigma({\bf p})$ and $n_{\sigma}({\bf
p})$ on $T$ and $B$. At FCQPT, eq. \eqref{HC1} can then be solved
analytically \cite{pr,ckz}. At $B=0$, the effective mass strongly
depends on $T$ demonstrating the NFL behavior \cite{pr,ckz}
\begin{equation}
M^*(T)\simeq a_TT^{-2/3}.\label{MTT}
\end{equation}
At finite $T$, the application of magnetic field $B$ drives the
system the to LFL region with
\begin{equation}
M^*(B)\simeq a_BB^{-2/3}.\label{MBB}
\end{equation}

A deeper insight into the behavior of $M^*(B,T)$ can be achieved
using some "internal" (or natural) scales. Namely, near FCQPT the
solutions of eq. \eqref{HC1} exhibit a behavior so that $M^*(B,T)$
reaches its maximum value $M^*_M$ at some temperature $T_{M}\propto
B$ \cite{pr}. It is convenient to introduce the internal scales
$M^*_M$ and $T_{M}$ to measure the effective mass and temperature.
Thus, we divide the effective mass $M^*$ and the temperature $T$ by
the values, $M^*_M$ and $T_{M}$, respectively. This generates the
normalized effective mass $M^*_N=M^*/M^*_M$ and the normalized
temperature $T_N=T/T_{M}$. Near FCQPT the normalized solution of
eq. \eqref{HC1} $M^*_N(T_N)$ can be well approximated by a simple
universal interpolating function \cite{pr}. The interpolation
occurs between the LFL and NFL regimes and represents the universal
scaling behavior of $M^*_N$ \cite{pr}
\begin{equation}M^*_N(y)\approx c_0\frac{1+c_1y^2}{1+c_2y^{8/3}}.
\label{UN2}
\end{equation}
Here, $y=T_N=T/T_{M}$, $c_0=(1+c_2)/(1+c_1)$, $c_1$, $c_2$ are
fitting parameters. Magnetic field $B$ enters eq. \eqref{HC1} only
in the combination $\mu_BB/T$, making $T_{M}\sim \mu_BB$. It
follows from eq.~\eqref{UN2} that
\begin{equation}
\label{TMB} T_M\simeq a_1\mu_BB,
\end{equation}
where $a_1$ is a dimensionless factor. Thus, in the presence of
fixed magnetic field the variable $y$ becomes $y=T/T_{M}\sim
T/\mu_BB$. Taking into account eq. \eqref{TMB}, we conclude that
eq. \eqref{UN2} describes the scaling behavior of the effective
mass as a function of $T$ versus $B$ - the curves $M^*_{N}$ at
different magnetic fields $B$ merge into a single one in terms of
the normalized variable $y=T/T_M$. Since the variables $T$ and $B$
enter symmetrically, eq. \eqref{UN2} describes the scaling behavior
of $M^*_{N}(B,T)$ as a function of $B$ versus $T$. The
normalization procedure deserves a remark here. Namely, since the
magnetic field dependence of $M^*_{N}(B,T)$ at fixed $T$ does not
have a maximum, the normalization is performed at its inflection
point, occurring at $B=B_{inf}$. As a result, we have $y=B/B_{inf}$
and $M_N^*=M^*(B,T)/M^*(B_{inf},T)$. In other words, the curves
$M^*_{N}$ at different $T$ merge into a single one in terms of the
normalized variable $y=B/B_{inf}$, while eq. \eqref{TMB} transforms
into the equation
\begin{equation}
\label{BMT} \mu_BB_{inf}\simeq a_2T,
\end{equation}
with $a_2$ is a dimensionless factor.

Now we construct the schematic phase diagram of SCQSL of the
organic insulators $\rm EtMe_3Sb[Pd(dmit)_2]_2$ and $\rm
\kappa-(BEDT-TTF)_2Cu_2(CN)_3$. The phase diagram is reported in
fig. \ref{fig0}. We assume that at $T=0$ and $B=0$ the system is
approximately located at FCQPT without tuning. Both magnetic field
$B$ and temperature $T$ play the role of the control parameters,
shifting the system from FCQPT and driving it from the NFL to LFL
regions as shown by the vertical and horizontal arrows. At fixed
temperatures the increase of $B$ drives the system along the
horizontal arrow from the NFL region to LFL one. On the contrary,
at fixed magnetic field and increasing temperatures the system
transits along the vertical arrow from the LFL region to the NFL
one. The hatched area denoting the transition region separates the
NFL state from the weakly polarized LFL state and contains the
solid line tracing the transition region, $T^*(B)\simeq T_M(B)$.
Referring to eq.~\eqref{TMB}, this line is defined by the function
$T^*\propto \mu_BB$, and the width $W(B)$ of the NFL state is seen
to be proportional $T$. In the same way, it can be shown that the
width $T^W(B)$ of the transition region is also proportional to
$T$.

\begin{figure} [! ht]
\begin{center}
\includegraphics [width=0.49\textwidth]{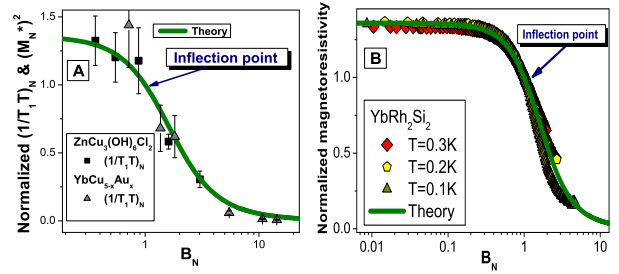}
\end{center}
\vspace*{-0.3cm} \caption{(Color online). Panel {\bf A}. The
relaxation properties of the herbertsmithite versus those of HF
metals. The normalized spin-lattice relaxation rate $(1/T_1T)_N$ at
fixed temperature as a function of magnetic field: Squares
correspond to data on $(1/T_1T)_N$ extracted from measurements on
$\rm ZnCu_3(OH)_6Cl_2$ \cite{imai}, while the triangles correspond
to those extracted from measurements on $\rm YbCu_{5-x}Au_{x}$ with
$x=0.4$ \cite{carr}. The inflection point, representing the
transition region, where the normalization is taken is shown by the
arrow. Our calculations based on eqs. \eqref{HC1} and eq.
\eqref{UN2} are depicted by the solid curve, tracing the scaling
behavior of $(M^*_N)^2$ and representing the $B$-dependence of the
thermal resistivity $w$, see main text and eq. \eqref{kap}. Panel
{\bf B}. The normalized longitudinal magnetoresistivity $\rho_N$
versus $B_N$, $\rho_N$ is extracted from measurements on $\rm
YbRh_2Si_2$ at different temperatures \cite{steg} listed in the
legend. The solid curve represents our calculations of
$(M^*_N)^2$.}\label{T1}
\end{figure}

As it was mentioned above, SCQSL plays a role of HF liquid. Thus,
we expect that SCQSL in organic insulators behaves like the
electronic HF liquid in HF metals, provided that the charge of an
electron were zero. In that case, the thermal resistivity $w$ of
SCQSL is related to the thermal conductivity $\kappa$
\begin{equation}\label{kap}
w=\frac{L_0T}{\kappa}=w_0+A_wT^2.
\end{equation}
In magnetic fields, the resistivity $w$ behaves like the electrical
magnetoresistivity $\rho_B=\rho_0+A_{\rho}T^2$ of the electronic
liquid, since $A_w$ represents the contribution of spinon-spinon
scattering to the thermal transport, being analogous to the
contribution $A_{\rho}$ to the charge transport, defined by
electron-electron scattering. Here, $L_0$ is the Lorenz number,
$\rho_0$ and $w_0$ are residual resistivity of electronic liquid
and QSL, respectively, and the coefficients $A_w\propto (M^*_{\rm
mag})^2$ and $A_{\rho}\propto (M^*)^2$ \cite{pr}. Thus, in the LFL
region the coefficient $A_w$ of the thermal resistivity of SCQSL
under the application of magnetic fields at fixed temperature
behaves like the spin-lattice relaxation rate shown in fig.
\ref{T1}, $A_w(B)\propto A_{\rho}\propto 1/T_1T(B)\propto
(M^*(B)_{\rm mag})^2$ \cite{epl}. In accordance with eq.
\eqref{MBB} as seen from fig. \ref{T1}, panel {\bf A}, the magnetic
field $B$ progressively reduces $1/T_1T$ \cite{imai,carr}, and
$1/T_1T$ as a function of $B$ possesses an inflection point at some
$B=B_{inf}$ shown by the arrow. The same behavior is seen from fig.
\ref{T1}, the panel {\bf B}: The magnetic field $B$ diminishes the
longitudinal magnetoresistivity \cite{steg}, and it as a function
of $B$ possesses an inflection point shown by the arrow. This
behavior is consistent with the phase diagram displayed in fig.
\ref{fig0}: At growing magnetic fields the NFL behavior first
converts into the transition one and then transforms into the LFL
behavior.

Figure \ref{T1}, panels {\bf A} and {\bf B}, display the normalized
spin-lattice relaxation rates $(1/T_1T)_N$ and the longitudinal
magnetoresistivity $\rho_B$ at fixed temperature versus normalized
magnetic field $B_N$, correspondingly. To clarify the universal
scaling behavior of the herbertsmithite  and the HF metal $\rm
YbCu_{5-x}Au_{x}$, we normalize both the functions $1/T_1T$ and
$(\rho_B-\rho_0)$ and the magnetic field. Namely, we normalize the
functions by their values at the inflection point, and magnetic
field is normalized by $B_{inf}$, $B_N=B/B_{inf}$. Since
$(1/T_1T)_N=\rho_B-\rho_0=(M^*_N(B))^2$ \cite{pr,epl}, we expect
that the different strongly correlated Fermi systems located near
FCQPT exhibit the same behavior of the effective mass, as it is
seen from fig. \ref{T1}, panels {\bf A} and {\bf B}. We shall see
below that the heat conductivity of the organic insulators exhibits
the same behavior.

Study of the thermal resistivity $w$ given by eq. \eqref{kap}
allows one to reveal spinons as itinerant excitations. It is
important that $w$ is not contaminated by contributions coming from
localized excitations. The temperature dependence of thermal
resistivity $w$ represented by the finite term $w_0$ directly shows
that the behavior of SCQSL is similar to that of metals, and there
is a finite residual term $\kappa/T$ in the zero-temperature limit
of $\kappa$. The presence of this term immediately proves that
there are gapless excitation associated with the property of normal
and HF metals, in which gapless electrons govern the heat and
charge transport, revealing a connection between the classical
physics and quantum criticality \cite{prb_T}. The finite $w_0$
means that in QSL both $k/T$ and $C_{mag}/T\propto M^*_{\rm mag}$
remain nonzero at $T\to0$. Therefore, gapless spinons, forming the
Fermi surface, govern the specific heat and the transport. Key
information on the nature of spinons is further provided by the
$B$-dependence of the coefficient $A_w$. The specific
$B$-dependence of $(1/T_1T)_N\propto(M^*_{\rm mag})^2$, shown in
fig. \ref{T1}, panel {\bf A}, and given by eq. \eqref{MBB},
establishes the behavior of QSL as SCQSL. We note that the heat
transport is polluted by the phonon contribution. On the other
hand, the phonon contribution is hardly influenced by the magnetic
field $B$. Therefore, we expect the $B$-dependence of the heat
conductivity to be governed by $A_w(B,T)$. Consider the approximate
relation,
\begin{eqnarray}
\nonumber 1&-&\frac{A_w(B,T)}{A_w(0,T)}=
1-\left(\frac{M^*(B,T)_{\rm mag}}{M^*(0,T)_{\rm mag}}\right)^2\\
&\simeq&a(T)\frac{\kappa(B,T)-\kappa(0,T)}{\kappa(0,T)}\equiv
a(T)I(B,T),\label{TR}
\end{eqnarray}
where the coefficient $a(T)$ is $B$-independent. To derive
\eqref{TR}, we employ eq. \eqref{kap}, and obtain
\begin{equation}\label{TR1}
\frac{\kappa}{L_0T}=\frac{1}{w_0+A_wT^2}+bT^2.
\end{equation}
Here, the term $bT^2$ describes the phonon contribution to the heat
transport. Upon carrying out simple algebra and assuming that
$[1-A_w(B,T)/A_w(0,T)]<1$, we arrive at eq. \eqref{TR}. It is seen
from fig. \ref{T1}, the both panels, that the effective mass
$M^*_N(B)\propto M^*_{\rm mag}(B)$ is a diminishing function of
magnetic field $B$. Then, it follows from eqs. \eqref{MBB},
\eqref{UN2} and \eqref{TR} that the function
$I(B,T)=[\kappa(B,T)-\kappa(0,T)]/\kappa(0,T)$ increases at elevated
field $B$ in the LFL region, while $I(B,T)\simeq 0$ in the NFL
region, for the function is approximately independent of $B$ in that
case.

\begin{figure} [! ht]
\begin{center}
\vspace*{-0.2cm}
\includegraphics [width=0.47\textwidth]{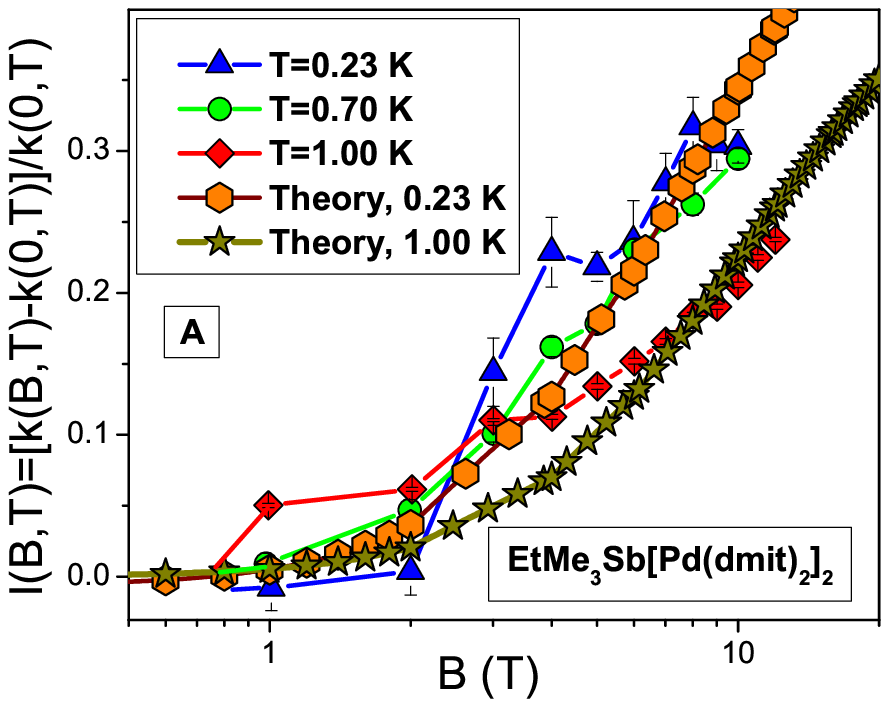}
\includegraphics [width=0.47\textwidth]{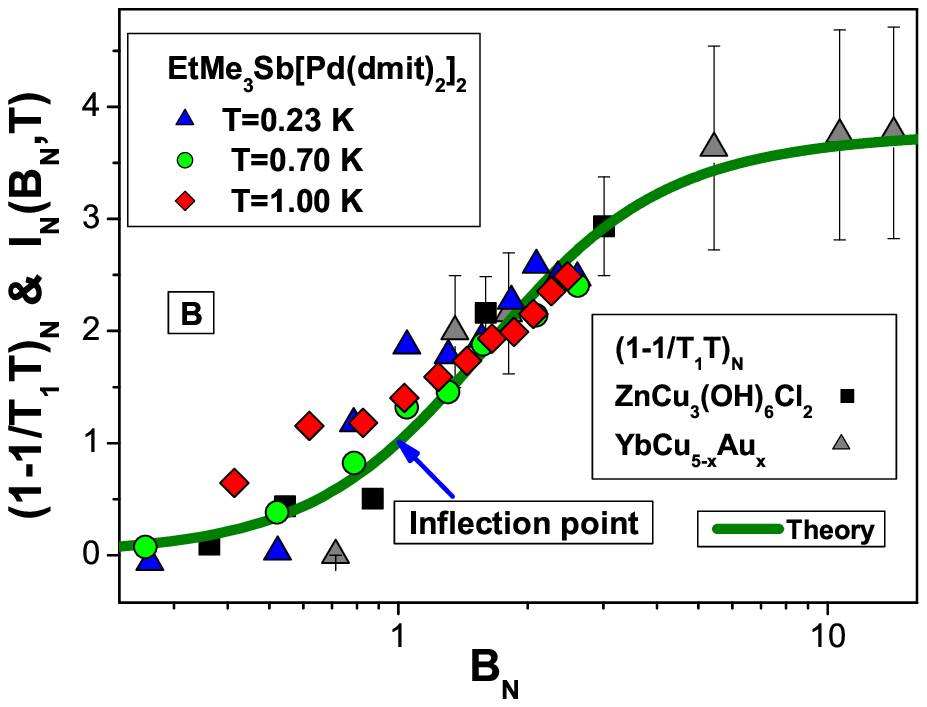}
\end{center}
\caption{(Color online). Panel {\bf A}: Magnetic field $B$
dependence of the thermal conductivity $I(B,T)$ measured on the
organic insulator $\rm EtMe{_3}Sb[Pd(dmit)_{2}]_{2}$ and
standardized by the zero field value $\kappa$,
$I(B,T)=[\kappa(B,T)-\kappa(B=0,T)]/\kappa(B=0,T)$ at temperatures
shown in the legend \cite{scqsl,chqs1}. Our calculations are based
on eq. \eqref{TR} and shown by pentagons and stars. Panel {\bf B}:
The normalized thermal conductivity $I_N(B_N,T)$ versus $B_N$ shown
by geometrical figures is extracted from the data shown in the
panel {\bf A} of this figure. The inflection point is shown by the
arrow. The magnetic field dependence of the function $(1-1/T_1T)_N$
is extracted from measurements of $(1/T_1T)_N$ shown in fig.
\ref{T1}, panel {\bf A}. The solid curve is obtained from the
theoretical curve in fig. \ref{T1}.} \label{kappa}
\end{figure}
\begin{figure} [! ht]
\begin{center}
\vspace*{-0.2cm}
\includegraphics [width=0.47\textwidth]{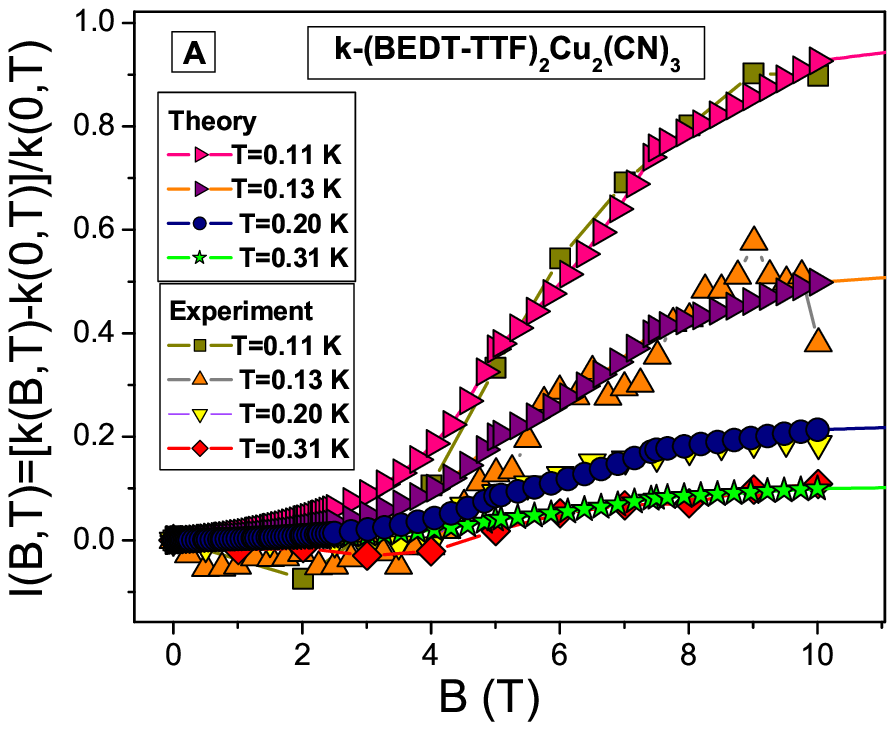}
\includegraphics [width=0.47\textwidth]{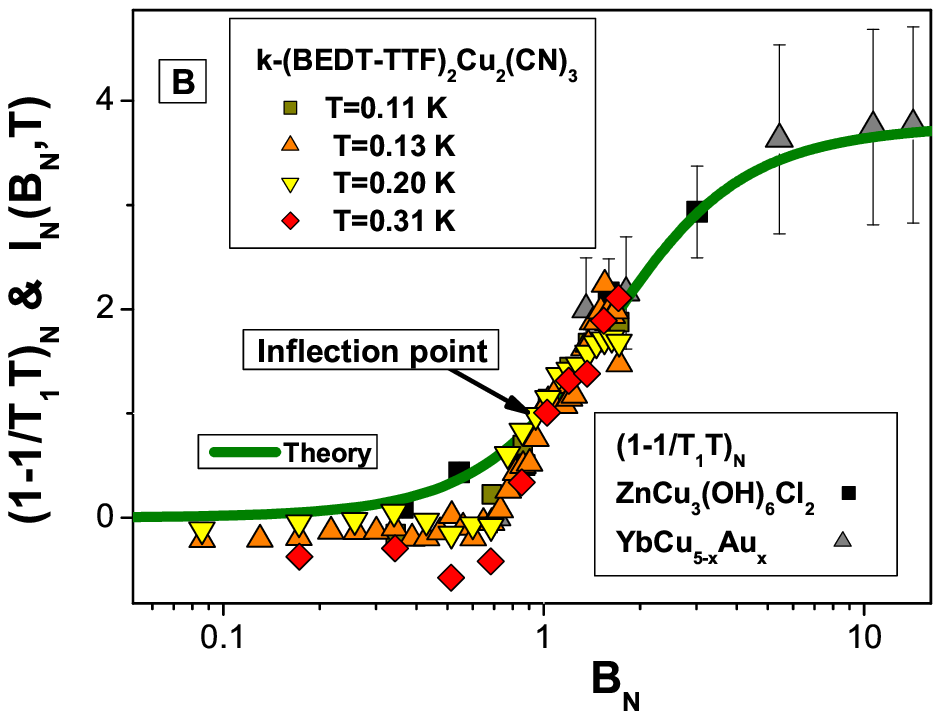}
\end{center}
\vspace*{-0.3cm} \caption{(Color online). Panel {\bf A}: Magnetic
field $B$ dependence of the thermal conductivity $I(B,T)$ measured
on the organic insulator $\rm\kappa-(BEDT-TTF)_2Cu_2(CN)_3$ and
standardized by the zero field value $\kappa$,
$I(B,T)=[\kappa(B,T)-\kappa(B=0,T)]/\kappa(B=0,T)$ at temperatures
shown in the legend \cite{chqs1}. Our calculations are based on eq.
\eqref{TR} and the results are shown by geometrical figures as
displayed in the legend. Panel {\bf B}: $I_N(B_N,T)$ versus $B_N$
shown by geometrical figures is extracted from the data shown in
the panel {\bf A}, of this figure. The magnetic field dependence of
the function $(1-1/T_1T)_N$ is extracted from measurements of
$(1/T_1T)_N$ shown in fig. \ref{T1}, panel {\bf A}. The solid curve
is obtained from the theoretical curve in fig.
\ref{T1}.}\label{kappa1}
\end{figure}

Recent measurements of $\kappa(B)$ on the organic insulators $\rm
EtMe{_3}Sb[Pd(dmit)_{2}]_{2}$ and
$\rm\kappa-(BEDT-TTF)_2Cu_2(CN)_3$ \cite{scqsl,chqs1} are displayed
in figs. \ref{kappa} and \ref{kappa1}, panels {\bf A}. The
measurements show that the heat is carried by phonons and SCQSL,
for the heat conductivity is well fitted by $\kappa/T=b_1+b_2T^2$,
where $b_1$ and $b_2$ are constants. The finite $b_1$ term implies
that spinon excitations are gapless in $\rm
EtMe{_3}Sb[Pd(dmit)_{2}]_{2}$, while in
$\rm\kappa-BEDT-TTF)_2Cu_2(CN)_3$ gapless excitations are under
debate \cite{chqs1}. A simple estimation indicates that the
ballistic propagation of spinons seems to be realized in the case
of $\rm EtMe{_3}Sb[Pd(dmit)_{2}]_{2}$ \cite{scqsl,chqs1}. It is
seen from figs. \ref{kappa} and \ref{kappa1}, panels {\bf A}, that
$I(B,T)=[\kappa(B,T)-\kappa(B=0,T)]/\kappa(B=0,T)$ demonstrates a
strong $B$-dependence, namely the field dependence shows an
increase of thermal conductivity for rising fields $B$. Such a
behavior is in agreement with eq. \eqref{MBB} and fig. \ref{T1}
which demonstrate that $(M^*_{mag}(B))^2$ is a diminishing function
of $B$. As a result, it follows from eq. \eqref{TR} that $I(B,T)$
is an increasing function of $B$. Our calculations based on eqs.
\eqref{epta} and \eqref{TR} are depicted by geometrical figures in
figs. \ref{kappa} and \ref{kappa1}, panels {\bf A}. Since we cannot
calculate the parameter $a(T)$ entering eq. \eqref{TR} we use it as
a fitting parameter. Temperature $T$ was also used to fit the data
at temperatures shown in the legend in figs. \ref{kappa} and
\ref{kappa1}. It is seen from figs. \ref{kappa} and \ref{kappa1},
panels {\bf A}, that $I(B,T)$ as a function of $B$ possesses an
inflection point at some $B=B_{inf}$. To reveal the scaling
behavior of the heat conductivity of the organic insulators, we
normalize both the function $I(B,T)$ and the magnetic field by
their values at the inflection points, as it was done in the case
of $(1/T_1T)$, see fig. \ref{T1}. The normalized heat conductivity
$I_N(B_N,T)$ does not depend on the factor $a(T)$, entering eq.
\eqref{TR}, and its calculations do not have any fitting
parameters. It is seen from figs. \ref{kappa} and \ref{kappa1},
panels {\bf B}, that in accordance with eq. \eqref{UN2}
$I_N(B_N,T)$ exhibits the scaling behavior and becomes a function
of a single variable $B_N$. It is instructive to compare the
normalized values of the function
$(1-1/T_1T)_N\equiv(1-[M^*(B,T)/M^*(B=0,T)]^2)_N$ extracted from
measurements of $(1/T_1T)_N$ shown in fig. \ref{T1}, panel {\bf A},
with $I_N(B_N,T)$. The extracted values are normalized by their
values at the inflection points and magnetic field is normalized by
$B_{inf}$, as it is done in the case of $(1/T_1T)_N$. It is seen
from figs. \ref{kappa} and \ref{kappa1}, panels {\bf B}, that
$(1-1/T_1T)_N$ and $I_N(B_N,T)$ are in good overall agreement with
the solid curve depicting the theoretical function
$(1-[M^*(B,T)/M^*(B=0,T)]^2)_N$, received from our calculations
represented by the solid curve in fig. \ref{T1}, the both panels.
It is seen that this function demonstrates a flat dependence at low
$B_N$, for at $B_N<1$ the system in its NFL state and the
$B$-dependence is weak. Thus, there is no need to introduce
additional quasiparticles activated by the application of magnetic
field in order to explain the growth of $I(B,T)$ at elevated $B$
\cite{scqsl,chqs1}. It is also seen from both figs. \ref{kappa} and
\ref{kappa1}, panels {\bf B}, the organic insulators demonstrate
the same behavior as $\rm ZnCu_3(OH)_6Cl_2$, $\rm
YbCu_{5-x}Au_{x}$, and $\rm YbRh_2Si_2$.

In summary, for the first time, we have explained magnetic
field-dependence of the low-temperature thermal conductivity
$\kappa$ in the organic insulators $\rm EtMe_3Sb[Pd(dmit)_2]_2$ and
$\rm \kappa-(BEDT-TTF)_2Cu_2(CN)_3$. Our analysis allows us to
detect SCQSL in these organic insulators, exhibiting the universal
scaling behavior.

This work was supported by U.S. DOE, Division of Chemical Sciences,
Office of Basic Energy Sciences, Office of Energy Research, AFOSR,
and U.S ARO (Grant W911NF-11-1-0194).


\end{document}